\documentclass[aps, prd, amsmath, floats, floatfix, twocolumn, nofootinbib, superscriptaddress, showpacs]{revtex4-1}
\usepackage{graphicx}
\usepackage{color}
\usepackage{latexsym}
\usepackage{amsfonts}

\newcommand{\be}{\begin{eqnarray}}
\newcommand{\ee}{\end{eqnarray}}

\begin{document}

\title{Iron K$\alpha$ line of Kerr black holes with Proca hair}

\author{Menglei Zhou}
\affiliation{Center for Field Theory and Particle Physics and Department of Physics, Fudan University, 200433 Shanghai, China}

\author{Cosimo Bambi}
\email[Corresponding author: ]{bambi@fudan.edu.cn}
\affiliation{Center for Field Theory and Particle Physics and Department of Physics, Fudan University, 200433 Shanghai, China}
\affiliation{Theoretical Astrophysics, Eberhard-Karls Universit\"at T\"ubingen, 72076 T\"ubingen, Germany}

\author{Carlos A. R. Herdeiro}
\affiliation{Departamento de F\`isica da Universidade de Aveiro and Center for Research and Development in Mathematics and Applications (CIDMA), Campus de Santiago, 3810-183 Aveiro, Portugal}

\author{Eugen Radu}
\affiliation{Departamento de F\`isica da Universidade de Aveiro and Center for Research and Development in Mathematics and Applications (CIDMA), Campus de Santiago, 3810-183 Aveiro, Portugal}

\date{\today}

\begin{abstract}
We continue our study on the capabilities of present and future X-ray missions to test the nature of astrophysical black hole candidates via X-ray reflection spectroscopy and distinguish Kerr black holes from other solutions of 4-dimensional Einstein's gravity in the presence of a matter field. Here we investigate the case of Kerr black holes with Proca hair~\cite{Herdeiro:2016tmi}. The analysis of a  sample of these configurations suggests that even extremely hairy black holes can mimic the iron line profile of the standard Kerr black holes, and, at least for the configurations of our study, we find that current X-ray missions cannot distinguish these objects from Kerr black holes. This contrasts with our previous findings for the case of Kerr black holes with scalar (rather than Proca) hair~\cite{work1}, even though such comparison may be biased by the limited sample. Future X-ray missions can detect the presence of Proca hair, but a theoretical knowledge of the expected intensity profile (currently missing) can be crucial to obtain strong constraints. 
\end{abstract}

\maketitle


\section{Introduction}

In Einstein's gravity, vacuum black holes (BHs) are described by the Kerr solution and are completely characterized by two parameters, associated, respectively, to the mass $M$ and the spin angular momentum $J$ of the object. This ``uniqueness theorem'' holds under specific assumptions~\cite{h1,h2,h3} (for a review, see e.g. Ref.~\cite{h4}). Besides vacuum (no matter energy-momentum tensor is allowed), the spacetime must have 3+1 dimensions and be stationary, asymptotically flat, regular on and outside the BH event horizon. The fact that BHs under such conditions can be completely specified by only two parameters, and not more, is often given the name of ``no-hair theorem''. This should be understood as the absence of other degrees of freedom ($e.g.$ multipolar structure) independent from the two aforementioned global charges, both of which are associated to Gauss laws and measurable at infinity. 

\bigskip

On the observational side, there is today a considerable body of evidence for the existence of BHs in our Galaxy and in the Universe. Stellar-mass BHs in X-ray binary systems are compact objects with a mass in the range 5-20~$M_\odot$, which is too high to be that of compact relativistic stars for any reasonable matter equation of state~\cite{o1,o2}. Supermassive BHs in galactic nuclei are too heavy, compact, and old to be clusters of non-luminous bodies, because the cluster lifetime due to evaporation and physical collision would be too short, thus making their existence in the present Universe highly unlikely~\cite{o3}. The non-detection of thermal radiation from the putative surface of all these objects is consistent with the fact that they are BHs with an event horizon and there is no surface~\cite{o4,o5}. The gravitational waves recently detected by the LIGO experiment are also consistent with the signal expected from the coalescence of stellar-mass BHs in Einstein's gravity~\cite{o6,o7,o8}.

\bigskip

The spacetime metric of astrophysical BHs formed from the complete gravitational collapse of very massive stars is thought to be well described by the Kerr solution. Initial deviations from the Kerr geometry are expected to be quickly radiated away with the emission of gravitational waves immediately after the creation of the event horizon~\cite{price}. Due to the difference between the proton and electron masses, the BH electric charge may be non-vanishing, but its equilibrium value is reached soon, because of the highly ionized host environment of these objects, and its impact on the background metric is completely negligible~\cite{ec}. The presence of an accretion disk can be ignored, because the density of the disk is low and its mass is several orders of magnitude smaller than the mass of the BH~\cite{disk1,disk2}. 

\bigskip

In this context, within General Relativity, a crucial assumption is that the Kerr metric is still the only physical BH solution even in the presence of matter~\cite{Berti:2015itd}. It turns out that even though the spirit of the uniqueness theorems can be extended, under assumptions, to different types of matter contents, yielding different no-hair theorems (see $e.g.$~\cite{Herdeiro:2015waa} for a review), it is possible to obtain BHs with ``hair" in Einstein's gravity. That is, BH solutions which are not fully described by parameters measurable at infinity and associated to Gauss laws (see~\cite{Herdeiro:2015waa,Sotiriou:2015pka,Volkov:2016ehx} for recent reviews). Whereas many of these examples have matter contents that violate the energy conditions, hairy BHs with a simple matter content and obeying all energy conditions have been recently discovered. These solutions are characterized by having \textit{synchronized hair}, in which the matter field has a phase angular velocity that matches the angular velocity of the rotating horizon. This synchronous rotation mechanism implies a vanishing matter field flux through the horizon and provides an equilibrium between the matter field and the horizon, preventing the collapse of the former into the latter.  Examples of solutions with synchronized hair include Kerr BHs with scalar hair~\cite{Herdeiro:2014goa,Herdeiro:2015gia,Herdeiro:2015tia} and Proca hair~\cite{Herdeiro:2016tmi}. The key-ingredient to circumvent classical no-hair theorems is that the bosonic field is complex and time periodic. The existence of hairy black hole solutions involving massive scalar fields has been proven rigorously in~\cite{Chodosh:2015oma}. These solutions interpolate continuously between vacuum Kerr BHs and a solitonic limit (scalar boson stars~\cite{Schunck:2003kk} and Proca stars~\cite{Brito:2015pxa}, respectively).  As such one expects their phenomenological properties to vary continuously between those of a standard Kerr BH and those of the corresponding solitons, which can potentially be very non-Kerr like.

\bigskip

There are two main approaches to test the nature of astrophysical BH candidates: with electromagnetic radiation~\cite{r-e} and with gravitational waves~\cite{r-gw}. In this paper, we continue our explorative study to understand the capabilities of present and future X-ray missions to test BH solutions in 4D Einstein's gravity in the presence of matter using X-ray reflection spectroscopy, the so-called iron line method~\cite{iron1,iron2,iron3,iron4}. We extend previous work on Kerr BHs with scalar hair~\cite{work1}, boson stars~\cite{work2}, and Proca stars~\cite{work3}.  Other phenomenological studies of BHs with synchronized hair include their shadows~\cite{Cunha:2015yba,Cunha:2016bpi,Vincent:2016sjq,Cunha:2016bjh}, Quasi-Periodic-Oscillations~\cite{Franchini:2016yvq} and some brief analyses of their quadrupoles and orbital frequency at the Innermost Stable Circular Orbit~\cite{Herdeiro:2014goa,Herdeiro:2015gia}.

\bigskip

The content of the present paper is as follows. In Sections~\ref{s-proca} and \ref{s-iron}, we briefly review, respectively, Kerr BHs with Proca hair and X-ray reflection spectroscopy. In Section~\ref{s-sim}, we present our simulations with the XIS instrument on board of the Suzaku X-ray mission (XIS/Suzaku) and with the LAD detector expected to be on board of eXTP (LAD/eXTP) and our best-fits. In Section~\ref{s-dis}, we discuss our results. Summary and conclusions are in Section~\ref{s-con}. Throughout the paper we employ natural units in which $c = G_{\rm N} = \hbar = 1$ and a metric with signature $(-+++)$.

\section{Kerr black holes with Proca hair \label{s-proca}}
Kerr BHs with Proca hair are solutions to Einstein's gravity minimally coupled to a complex Proca field~\cite{Herdeiro:2016tmi} and may represent the final product of the non-linear evolution of super-radiant instability~\cite{East:2017ovw}. The corresponding action is 
\begin{equation}
\mathcal{S}=\int d^4x \sqrt{-g}\left(\frac{R}{16 \pi  } -\mathcal{F}_{\alpha \beta}\bar{\mathcal{F}}^{\alpha\beta}-\frac{\mu^2}{2}\mathcal{A}_\alpha \bar{\mathcal{A}}^\alpha\right) \ ,
\end{equation}
where $\mathcal{A}$ is the complex Proca potential 1-form with field strength $\mathcal{F}=d\mathcal{A}$ and complex conjugates  $\bar{\mathcal{A}}$ and $\bar{\mathcal{F}}$, respectively; $\mu$ is the Proca field mass.

BH solutions to this model can be found using the following metric ansatz~\cite{Herdeiro:2016tmi} 
\begin{eqnarray}
 ds^2= && -e^{2F_0(r,\theta)}N(r) dt^2+e^{2F_1(r,\theta)}\left(\frac{dr^2}{N(r)}+r^2 d\theta^2\right) \nonumber \\
&& + e^{2F_2(r,\theta)}r^2 \sin^2\theta \left[d\varphi-W(r,\theta) dt\right]^2 \ ,
 \label{kerrnc}
\end{eqnarray}
where 
\begin{equation}
N(r) \equiv 1 -\frac{r_H}{r} \ ,
\label{n}
\end{equation} 
and $r_H$ is the horizon radial coordinate; the Proca potential ansatz is~\cite{Herdeiro:2016tmi}
\begin{eqnarray}
\mathcal{A}= e^{i(m\varphi-w t)}  &&  \left[
  iV(r,\theta) dt  +H_1(r,\theta)dr+H_2(r,\theta) d\theta \nonumber \right. \\
 && \left.    +i H_3(r,\theta) \sin \theta d\varphi 
\right] \ .
\label{procaclouds}
\end{eqnarray}
All of the above functions are real functions. 
Solutions exist under the synchronization condition 
\begin{equation}
w=m\Omega_H \ ,
\label{sync}
\end{equation}
where $\Omega_H=W(r=r_H,\theta)$ is the ($\theta$-independent) horizon angular velocity.

\begin{figure}[h!]
\begin{center}
\includegraphics[type=pdf,ext=.pdf,read=.pdf,width=8.0cm]{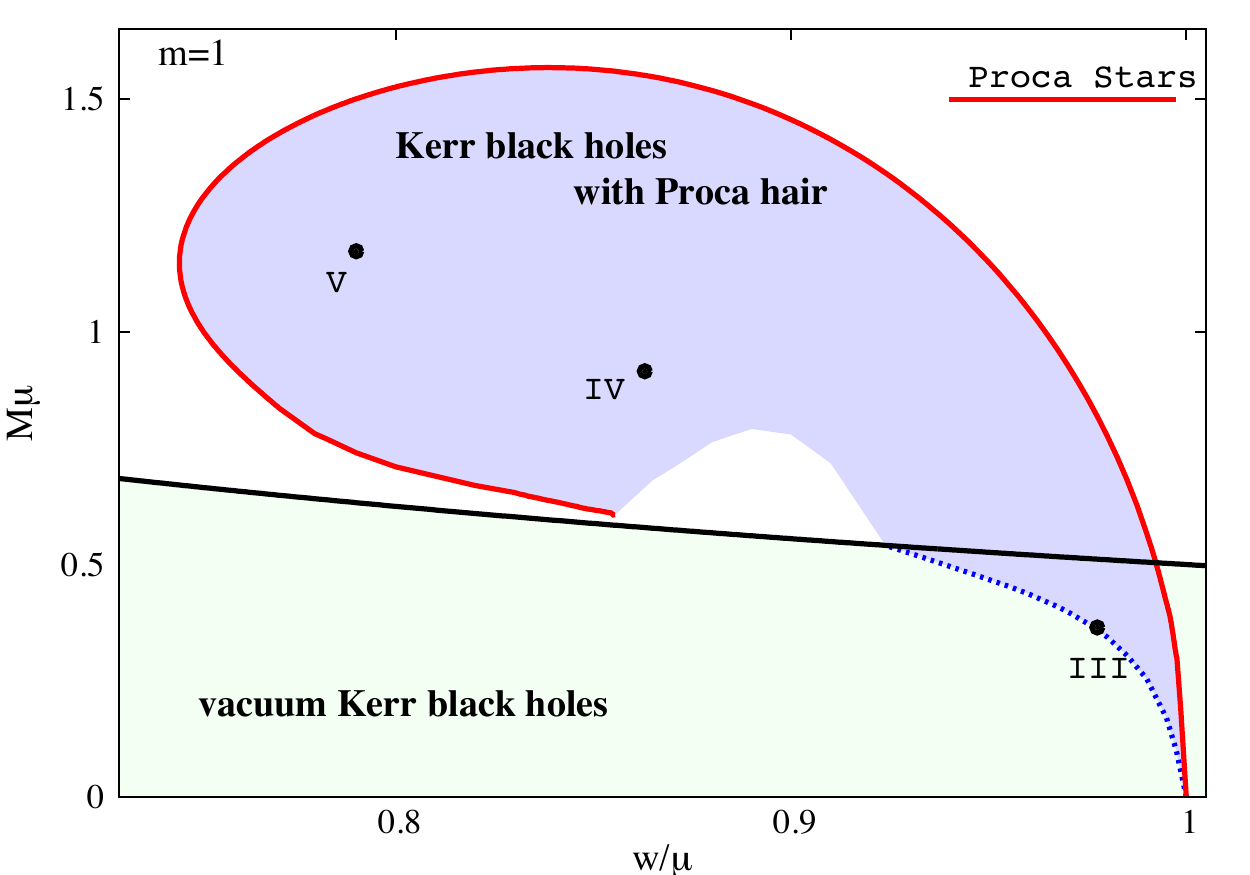}
\end{center}
\vspace{-0.5cm}
\caption{Domain of existence of Kerr BHs with Proca hair, in an ADM mass $vs.$ vector field frequency diagram in units of the Proca field mass $\mu$.
\label{f-proca}}
\end{figure}

In Fig.~\ref{f-proca} we exhibit the domain of existence for Kerr BHs with Proca hair in an ADM mass $vs.$ vector frequency diagram, both in units of the Proca field mass $\mu$. This domain of existence is for solutions with $m=1$ and also with the smallest number of nodes of the Proca field potential.  The two plotted boundaries for this domain of existence correspond to the solitonic limit (rotating Proca stars -- solid red line) and the Kerr limit (blue dotted line), corresponding to the subset of vacuum Kerr solutions that can support a Proca stationary cloud ($i.e$ a test field configuration on the Kerr background) with $m=1$. The BH solid line corresponds to the set of extremal Kerr BHs in such a diagram, translating $w$ into $\Omega_H$ via \eqref{sync}. Vacuum Kerr BHs exist below the black solid line; thus the vacuum Kerr limit of this family of hairy BHs (blue dotted line) is a 1-dimensional subset of the two dimensional parameter space. In (a subset of) the white region just above the black solid line around $\omega/\mu \sim 0.9$, there may still exist hairy BHs, in particular close to the border of the blue shaded area. However, this region is particularly challenging numerically, and no accurate solutions could be obtained.

 In Fig.~\ref{f-proca} there are three highlighted configurations, dubbed III, IV and V (to match their numbering in~\cite{Herdeiro:2016tmi}),  corresponding to the solutions that shall be analyzed below. These can be briefly described as a BH with ``little hair'' (only a relatively small fraction of the total mass and angular momentum is stored in the Proca field and the metric is thus fairly close to Kerr solution -- configuration III), a ``very hairy'' BH (most of its mass and angular momentum are stored in the Proca field -- configuration IV) and an ``extremely hairy'' BH (the mass and the angular momentum are almost fully in the Proca field -- configuration V). Their main physical properties are summarized in table~\ref{tab1} (adapted from~\cite{Herdeiro:2016tmi}).

\begin{table*}[t]
\begin{center}
\begin{tabular}{|c||c|c|c|c|c|c|c|c|c|}
\hline
Config. & $w/\mu$  & $r_H/\mu$ & $\mu M_{ADM}$ & $\mu^2J_{ADM}$ & $a_*^{ADM}$ & $\mu M_{H}$ & $\mu^2J_{H}$ & $\mu M^{(\mathcal{P})}$ & $\mu^2J^{(\mathcal{P})}$   \\ \hline\hline
III  & 0.98 & 0.25 & 0.365 & 0.128 & 0.961 & 0.354 & 0.117 & 0.011 & 0.011
\\ \hline
IV  & 0.86& 0.09 & 0.915 & 0.732 & 0.874 & 0.164 & 0.070 & 0.751 & 0.663
\\ \hline
V  & 0.79& 0.06 & 1.173 & 1.079 & 0.784 & 0.035 & 0.006 & 1.138 & 1.073
\\ \hline 
\end{tabular}\\ \bigskip
\caption{Configurations to be examined in this paper. Here $M_{ADM}$ and $J_{ADM}$ denote the ADM mass and angular momentum, $M_H$ and $J_H$ denote the horizon mass and angular momentum, and $M^{(\mathcal{P})}$ and $J^{(\mathcal{P})}$ denote the energy and angular momentum stored in the Proca field. The dimensionless spin parameter is $a_*^{ADM} = J_{ADM}/M_{ADM}^2$. Details on how these quantities are computed may be found in~\cite{Herdeiro:2016tmi}. Observe that for configurations III, IV and V, the percentages of energy and angular momentum stored inside the horizon are, respectively $\simeq$ (97\%, 91\%), (18\%, 9.5\%) and (3.0\%, 0.6\%). This justifies designating the three BHs as having little hair, being very hairy and being extremely hairy. The data is publicly available in~\cite{datakbhph}.}
\end{center}
\label{tab1}
\end{table*}

\section{X-ray reflection spectrum \label{s-iron}}

X-ray reflection spectroscopy is a technique that can be employed to probe the spacetime metric in the strong gravity region around a BH by studying the reflection spectrum of its accretion disk. Within the disk-corona model~\cite{corona1,corona2}, a BH is surrounded by a geometrically thin and optically thick accretion disk. The disk emits like a blackbody locally and as a multi-color blackbody when integrated radially. The ``corona'' is a hot, usually optically thin, cloud of gas close to the BH, but its exact geometry is currently unknown. For instance, it may be the base of the jet, a sort of atmosphere above the accretion disk, or some accretion flow between the disk and the BH. Due to inverse Compton scattering of the thermal photons from the disk off the hot electrons in the corona, the latter becomes an X-ray source with a power-law spectrum. Some X-ray photons from the corona illuminate the disk, producing a reflection component with some emission lines. The most prominent feature is usually the iron K$\alpha$ line, which is at about 6.4~keV for neutral and weakly ionized atoms but it can shift up to 6.97~keV in the case of H-like iron ions.

The accretion disk can be described by the Novikov-Thorne model~\cite{nt-model}, which is the standard set-up for geometrically thin and optically thick accretion disk. The disk is in the equatorial plane perpendicular to the spin of the compact object. The particles of the disk follow nearly geodesic circular orbits in the equatorial plane. The inner edge of the disk is located at the innermost stable circular orbit (ISCO). When the particles of the gas reach the ISCO, they quickly plunge onto the central object, so we can neglect the radiation emitted inside the ISCO.

The iron K$\alpha$ line is a very narrow emission line in the rest frame of the emitting gas, while the line in the reflection spectrum of BHs is broad and skewed, as a result of special and general relativistic effects occurring in the strong gravity region around the compact object. If properly understood and in the presence of high quality data, the study of the disk reflection spectrum can be a powerful tool to probe the geometry around BHs and check whether the spacetime metric is indeed well described by the Kerr solution~\cite{ss,jp,i1,i2,i3,i4}.

As an explorative work, for the sake of simplicity here we do not consider the full reflection spectrum and we restrict our attention to the iron K$\alpha$ line only, which is the most prominent feature and the most sensitive one to the strong gravity region. The calculations of the iron line profile as detected by an instrument in the flat faraway region have been extensively discussed in the literature, and in our case we employ the code described in Refs.~\cite{code1,code2} and readapted to treat numerical metrics in~\cite{menglei}. This is essentially a ray-tracing code. We fire photons from a grid in the image plane of the distant observer to the accretion disk and we find the emission points. We calculate the redshift factor from the photon constants of motion and the gas velocity at the emission point. We repeat the calculations for every point in the grid and we then integrate over the disk image to get the total spectrum as seen by the distant observer.

The shape of the iron line depends on the background metric (which determines all the relativistic effects: gravitational redshift, Doppler boosting, light bending), the inclination angle of the disk with respect to the line of sight of the distant observer, the emissivity profile of the accretion disk. The latter is a weak point in the calculations, because it should depend on the geometry of the corona, which is currently unknown. It is common to parametrize the intensity profile for a generic corona geometry with a power-law (the intensity is proportional to $1/r^q$, where $r$ is the emission radius and $q$ is the emissivity index) or with a broken power-law (namely $1/r^{q_1}$ for $r < r_{\rm br}$ and $1/r^{q_2}$ for $r > r_{\rm br}$, where $q_1$ and $q_2$ are, respectively, the inner and the outer emissivity indices and $r_{\rm br}$ is the breaking radius).

\begin{widetext}

\begin{figure*}[h!]
\begin{center}
\includegraphics[type=pdf,ext=.pdf,read=.pdf,width=9cm]{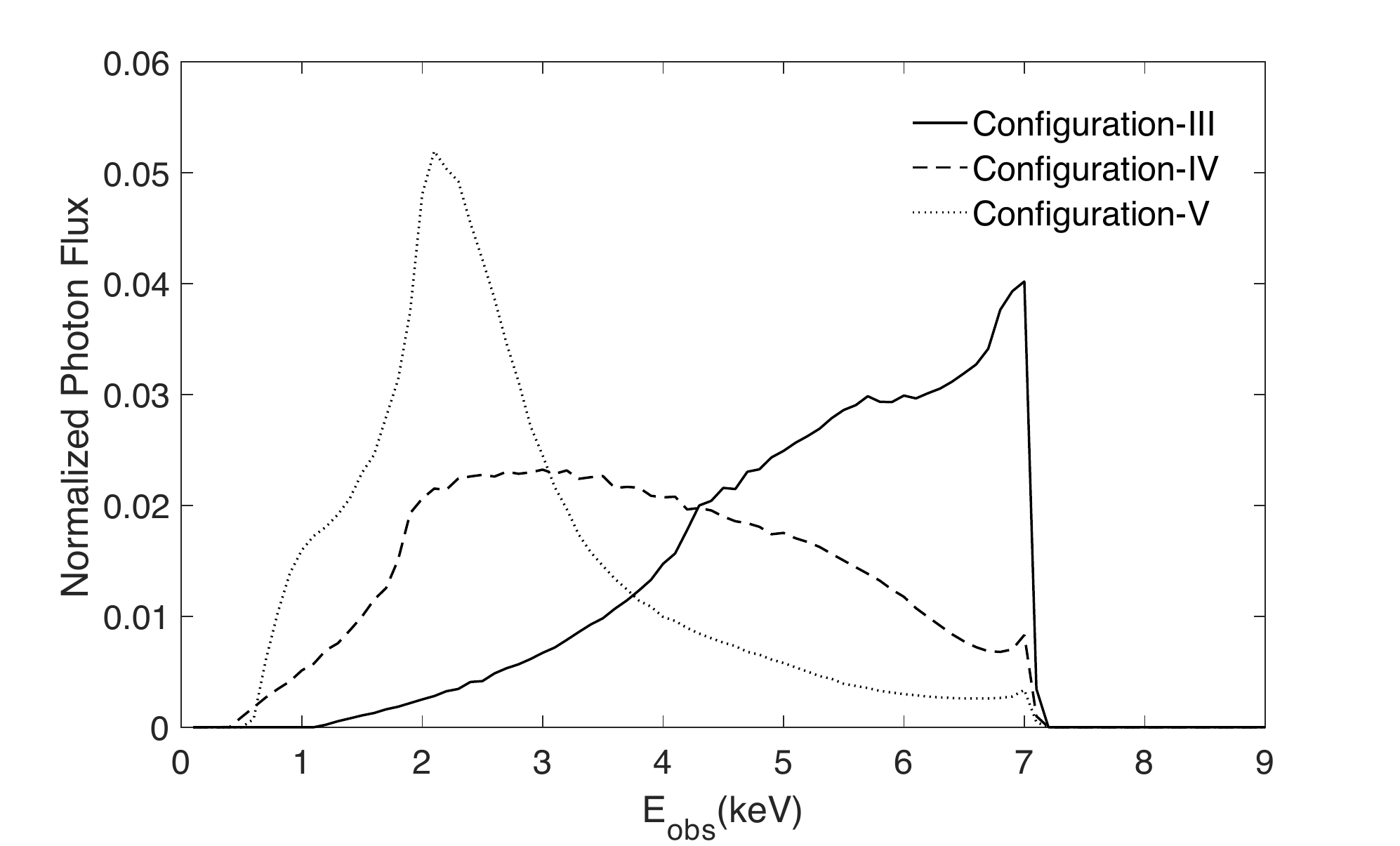}
\hspace{-0.5cm}
\includegraphics[type=pdf,ext=.pdf,read=.pdf,width=9cm]{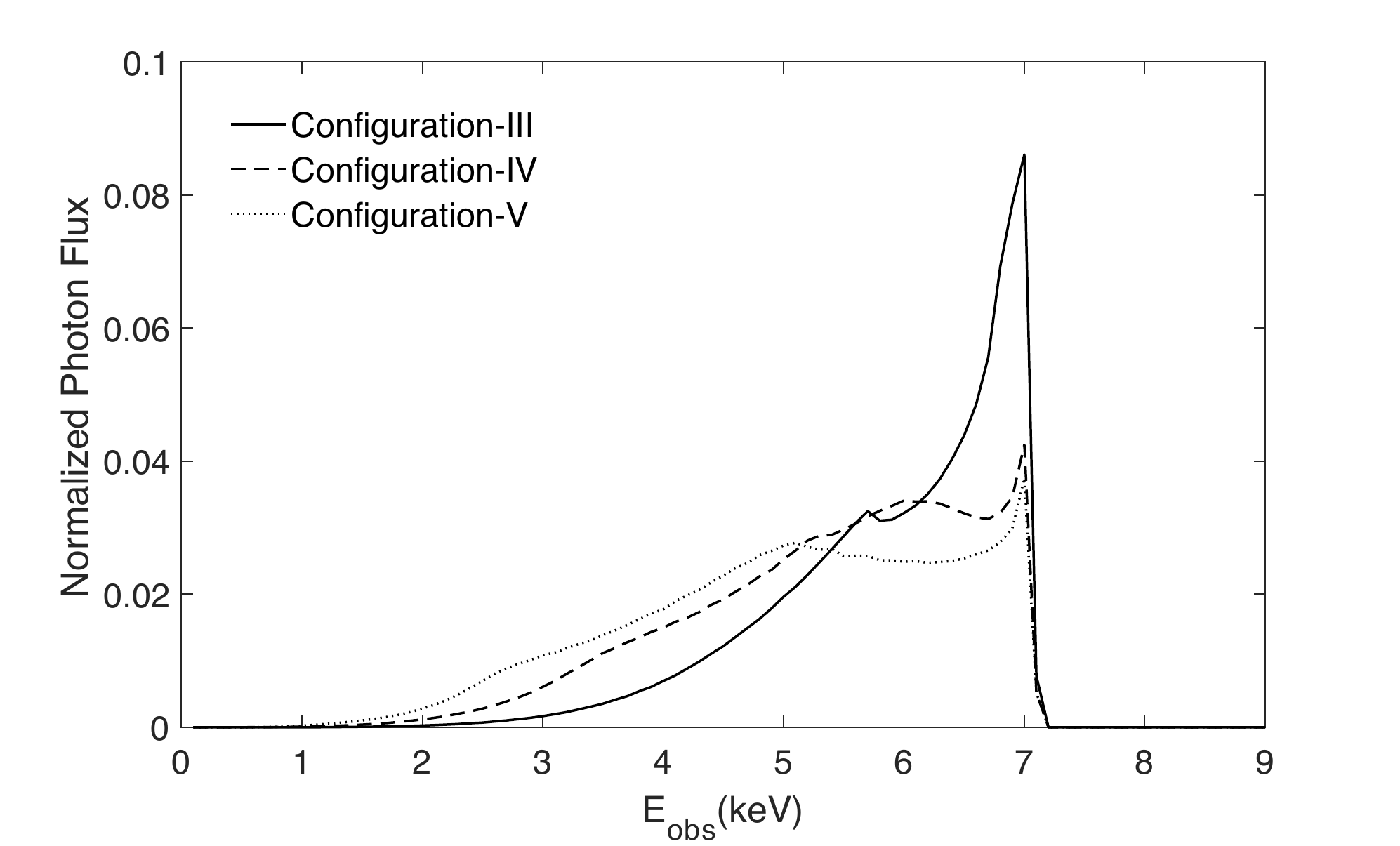}
\end{center}
\vspace{-0.3cm}
\caption{Iron line shapes of Kerr BHs with Proca hair in the configurations~III (solid lines), IV (dashed lines), and V (dotted lines) assuming the intensity profile $\propto 1/r^3$ (left panel) and $\propto h/(h^2 + r^2)^{3/2}$ with $h=2$ and $r$ and $h$ in units of the Proca field mass $\mu$
(right panel). The viewing angle is $i = 45^\circ$ and the energy of the line in the rest frame of the emitting gas is at 6.4~keV. The photon flux is normalized in such a way that if we integrate over the energy bins we obtain 1.
\label{f-iron}}
\end{figure*}

\end{widetext}

Fig.~\ref{f-iron} shows the iron line profiles for the configurations~III, IV, and V of Kerr BHs with Proca hair. In both panels, the inclination angle of the disk with respect to the line of sight of the distant observer is $i = 45^\circ$ and the rest-frame energy of the line is at 6.4~keV. In the left panel, it was assumed the intensity profile $1/r^3$. In the right panel, we assumed the lamppost-inspired intensity profile
\be
I \propto \frac{h}{(h^2 + r^2)^{3/2}} \, .
\ee 
It corresponds to the intensity profile expected in the Newtonian theory (no light bending) in the case the corona is a point-like source at the height $h$ from the central object and along its spin axis~\cite{dauser}. $h$ of order of a few gravitational radii is what we can expect if, for instance, the corona is the base of the jet.

As it is evident from Fig.~\ref{f-iron}, the shape of the iron line strongly depends on the emissivity profile. In particular, if we assume the intensity profile $1/r^3$, the solutions~IV and V have an iron line with a high photon count at low energies, which is not the case for the iron lines of Kerr BHs in vacuum. This happens because in the case of a Kerr BH with Proca hair the photons can be emitted closer to the center (so they can be significantly gravitationally redshifted) and can still escape to infinity. If we change the intensity profile, we cannot change the maximum and minimum energies of the photons reaching the distant observer (only determined by the background metric), but we can change the shape of the line. In this way we can decrease the photon count at low energies, as shown in the right panel in Fig.~\ref{f-iron}. As already stressed in previous work, see in particular Ref.~\cite{work3}, accurate tests of the metric around astrophysical BHs will only be possible in the presence of the correct astrophysical model for the corona. Phenomenological metrics, like a power-law or a broken power-law, are approximations unsuitable to get strong and reliable constraints on the spacetime metric with the high quality data expected with the next generation of X-ray facilities.

\section{Simulations \label{s-sim}}

We simulate observations with XIS/Suzaku\footnote{http://heasarc.gsfc.nasa.gov/docs/suzaku/}, as an example for studying the capabilities of current X-ray missions, and LAD/eXTP\footnote{http://www.isdc.unige.ch/extp/}~\cite{extp}, in order to explore the opportunities offered by the next generation of X-ray satellites. We employ the strategy already used in our studies of Kerr BHs with scalar hair~\cite{work1}, boson stars~\cite{work2}, and Proca stars~\cite{work3}, as well as to the study of other metrics~\cite{p1,p2}. We simulate observations of our metrics and we then fit the data with a Kerr model. If we find a good fit, this means that we cannot distinguish our model from a Kerr spacetime. If the fit is not acceptable, the model can potentially be tested with the mission in consideration. Current X-ray data of BHs are consistent with the Kerr metric, so a non-acceptable fit from XIS/Suzaku should rule out the spacetime. A non-acceptable fit with LAD/eXTP can instead be interpreted as the fact that such a mission can test that configuration. Note that this is a simple study, but we believe it can catch the key results and establish a proof of concept.

We use Xspec\footnote{Xspec is a package for X-ray data analysis. More details can be found at https://heasarc.gsfc.nasa.gov/xanadu/xspec/} with the redistribution matrix, ancillary, and background files of XIS/Suzaku and LAD/eXTP following the forward-folding approach common in X-ray astronomy. The spectrum measured by an instrument is given as a photon count per spectral bin and can be written as
\be\label{eq-ffa}
C(h) = \tau \int R(h,E) \, A(E) \, s(E) \, dE \, ,
\ee
where $h$ is the spectral channel, $\tau$ is the exposure time, $E$ is the photon energy, $R(h,E)$ is the redistribution matrix (essentially the response of the instrument), $A(E)$ is the effective area (say the efficiency of the instrument, which is given in the ancillary file), and $s(E)$ is the intrinsic spectrum of the source. In general, the redistribution matrix cannot be inverted, and for this reason we have to deal with $C(h)$ (folded spectrum). With the forward-folding approach, we consider a set of parameter values for the intrinsic spectrum, we find $C(h)$, we compare the ``observed spectrum'' with the folded spectrum with some goodness-of-fit statistical test (e.g. $\chi^2$), and we repeat all these steps to find ``the best fit'' by changing the input parameters in the theoretical model.

Our simulations are done assuming the intrinsic spectra $s(E)$ calculated in the Kerr spacetimes with Proca hair in the previous section and we add the background. The latter includes the noises of the instrument and of the environment (e.g. photons not from the target source or cosmic rays). The observational error has also the intrinsic noise of the source (Poisson noise) due to the fact the spectrum is as a photon count per bin and is not a continuous quantity. The main advantage of LAD/eXTP with respect to current X-ray instruments is the much larger effective area of the detector. The effective area of LAD/eXTP at 6~keV is about 35,000~cm$^2$. The effective area of XIS/Suzaku at 6~keV is about 250~cm$^2$. This means that, for the same source and the same exposure time, the photon count at 6~keV of LAD/eXTP is $35,000/250 \approx 150$ times the photon count of XIS/Suzaku. This significantly reduces the Poisson noise and leads a more precise measurement of the spectrum.

For these simulations, we do not consider a specific source, but we employ typical parameters for a bright BH binary, namely an X-ray binary with a stellar-mass BH and a stellar companion. This is likely the most suitable candidate for this kind of tests, because they are brighter than supermassive BHs and therefore we can get a more precise measurement of the spectrum. We assume that the energy flux in the 1-10~keV range is $10^{-9}$~erg/s/cm$^2$. The spectrum of the source is a power-law with photon index $\Gamma = 1.6$ (the spectrum is $\propto E^{-\Gamma}$) to describe the primary component from the corona and a single iron line with an equivalent width of 200~eV. The iron lines used in these simulations are those shown in Fig.~\ref{f-iron}, so we have six models, namely three different spacetimes (configurations~III, IV, and V) and two types of intensity profiles. The inclination angle of the disk is $45^\circ$. For both missions, we assume an exposure time of 100~ks.

The simulations are then treated as real data. After rebinning to ensure a minimum count rate per bin of 20 in order to use the $\chi^2$ statistics, we fit the data with a power-law and an iron line for a Kerr spacetime. For the latter, we use RELLINE~\cite{relline}, which is an Xspec model to describe a single relativistically broadened emission line from the accretion disk of a Kerr BH. We have 9~free parameters: the photon index of the power law $\Gamma$, the normalization of the power-law, inner and outer emissivity indices $q_1$ and $q_2$, the breaking radius $r_{\rm br}$, the spin parameter $a_*$, the inclination angle of the disk $i$, the outer radius of the disk, and the normalization of the iron line.

\begin{figure*}[h!]
\begin{center}
\includegraphics[type=pdf,ext=.pdf,read=.pdf,width=9cm]{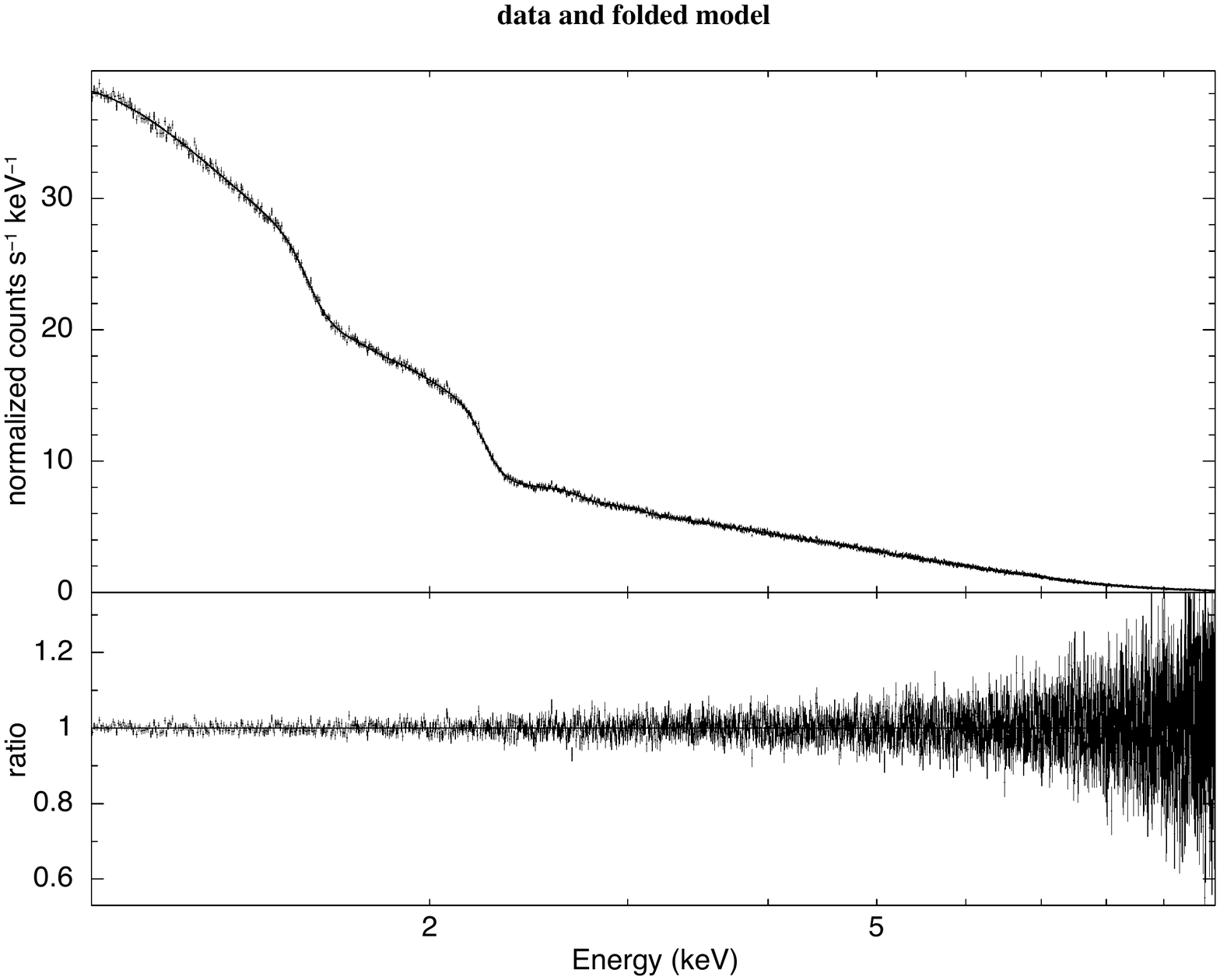}
\hspace{-0.5cm}
\includegraphics[type=pdf,ext=.pdf,read=.pdf,width=9cm]{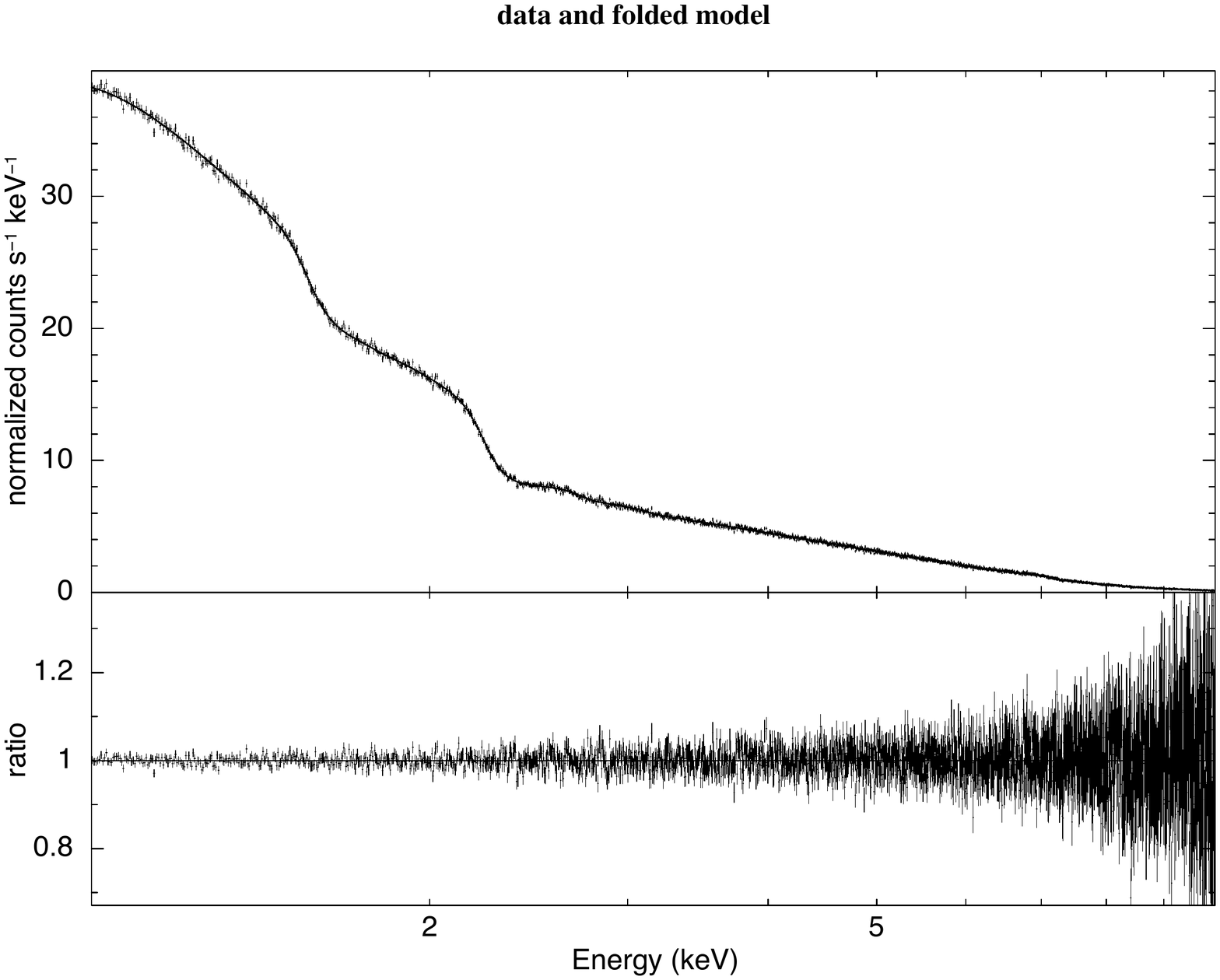} \\
\includegraphics[type=pdf,ext=.pdf,read=.pdf,width=9cm]{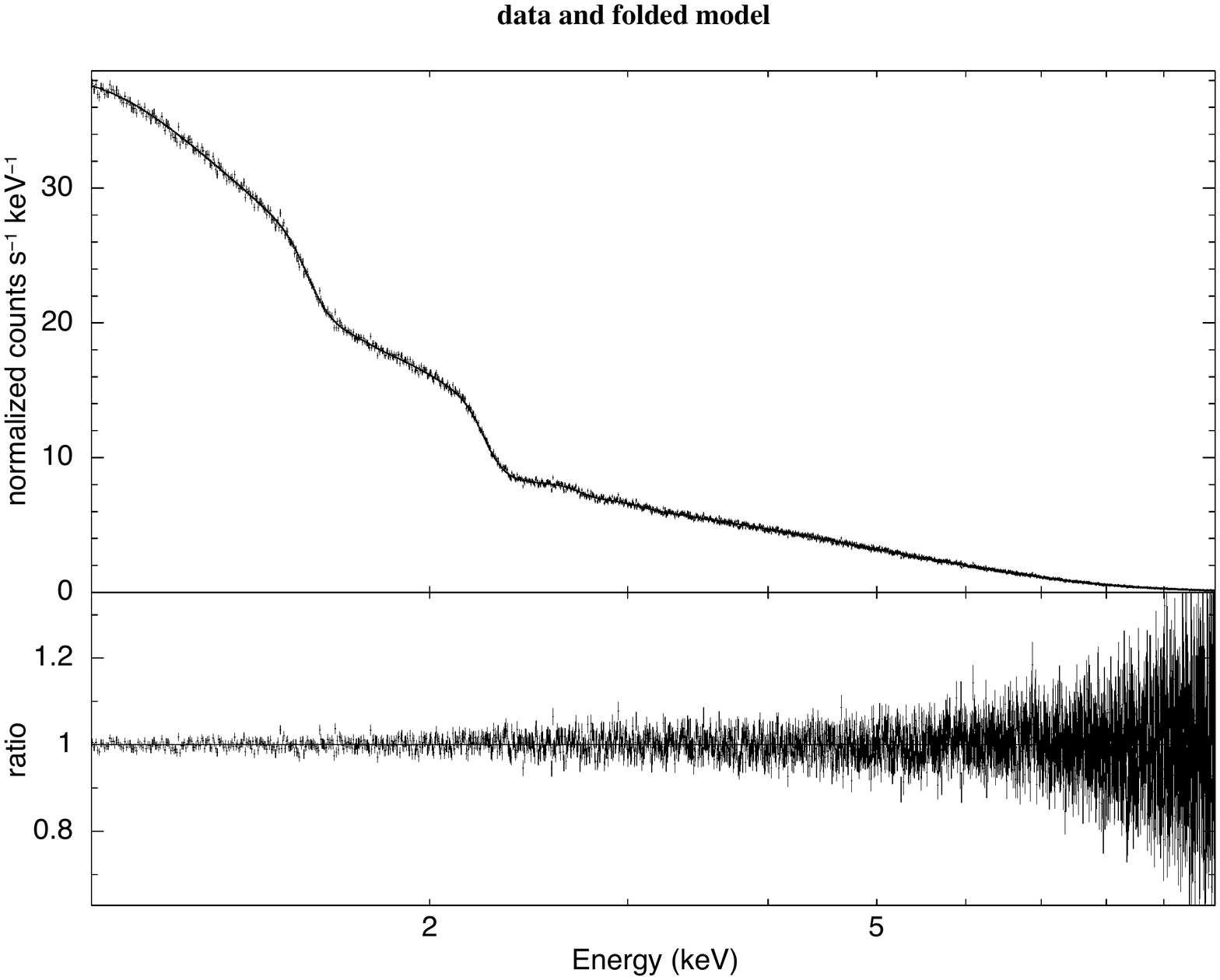}
\hspace{-0.5cm}
\includegraphics[type=pdf,ext=.pdf,read=.pdf,width=9cm]{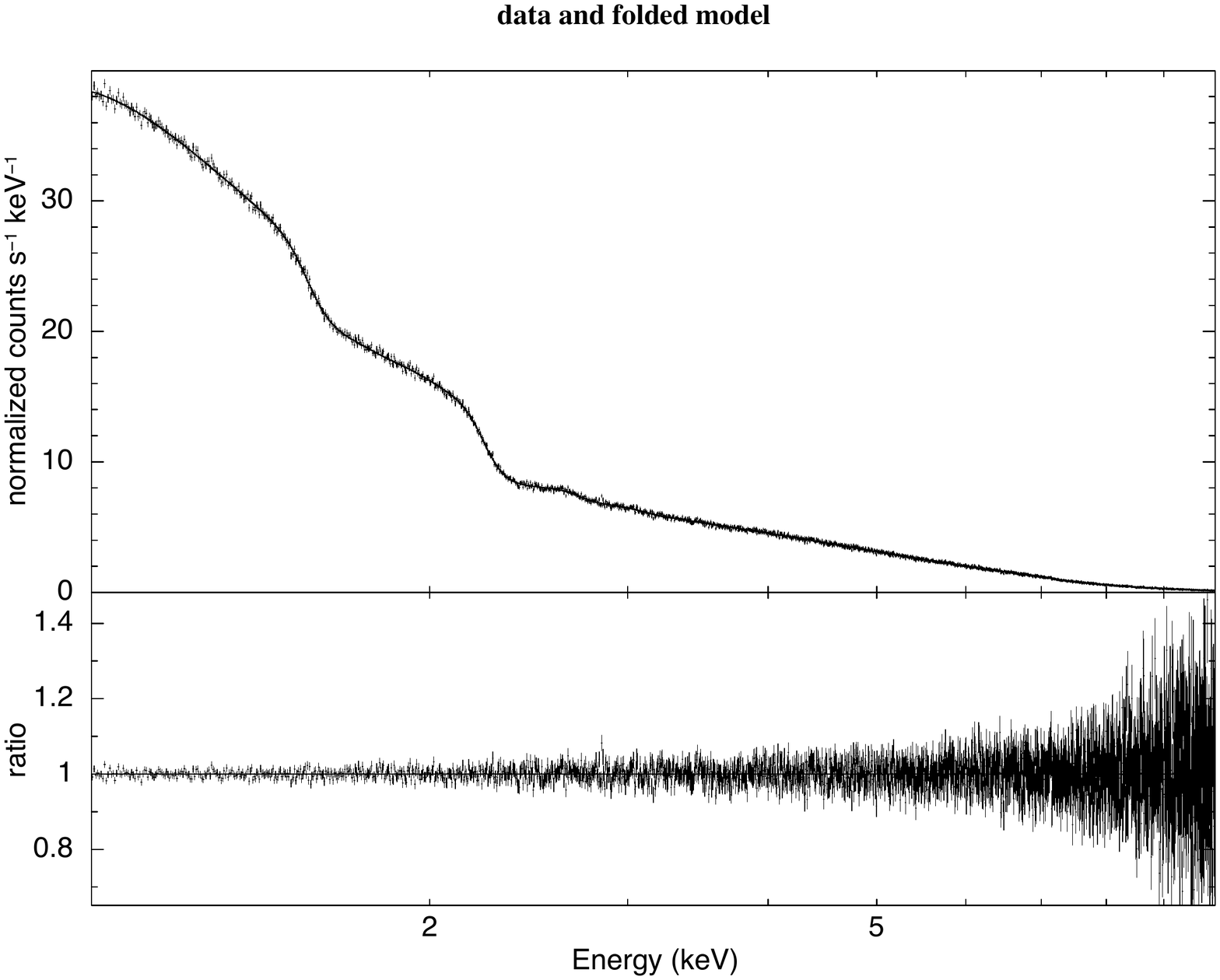} \\
\includegraphics[type=pdf,ext=.pdf,read=.pdf,width=9cm]{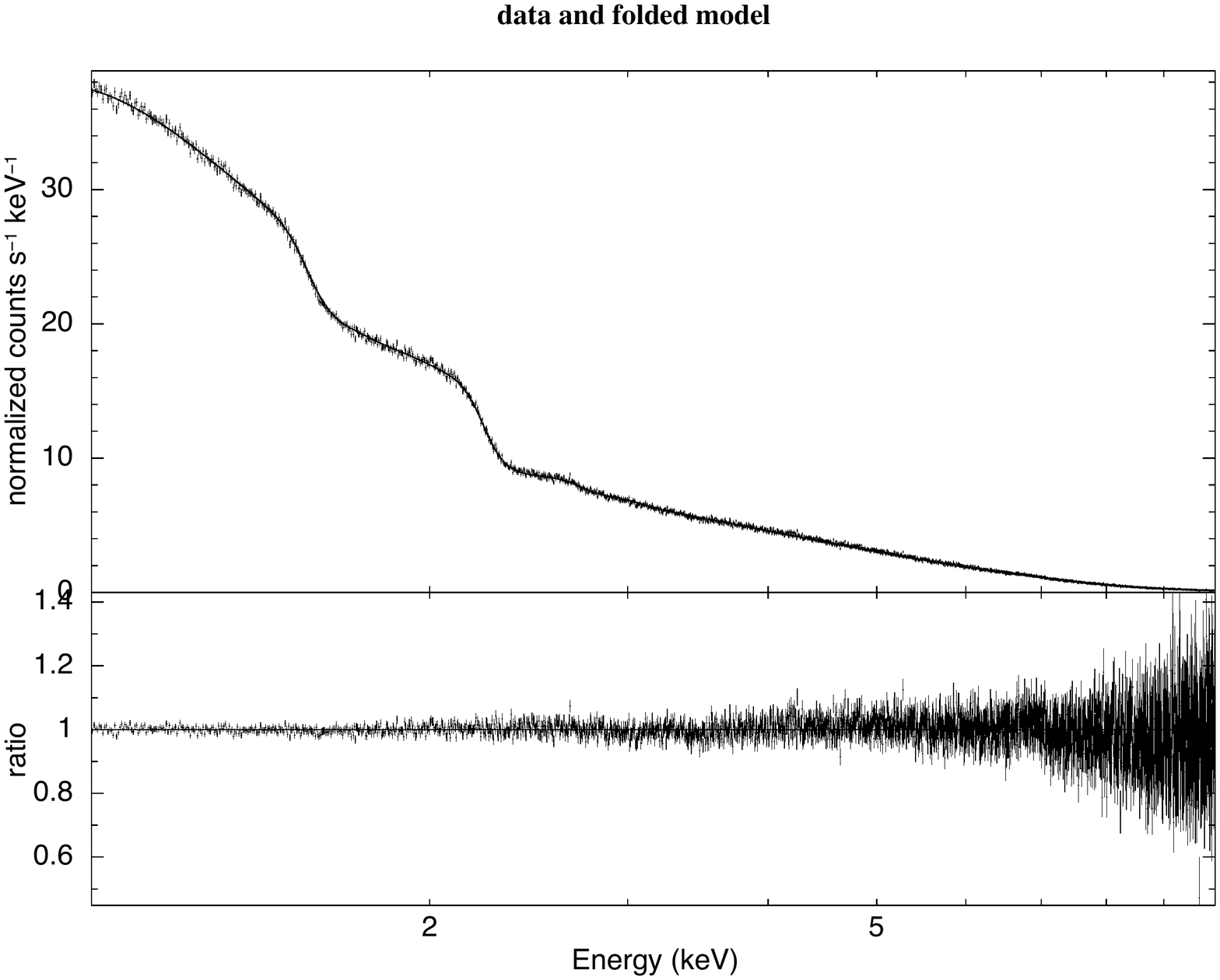}
\hspace{-0.5cm}
\includegraphics[type=pdf,ext=.pdf,read=.pdf,width=9cm]{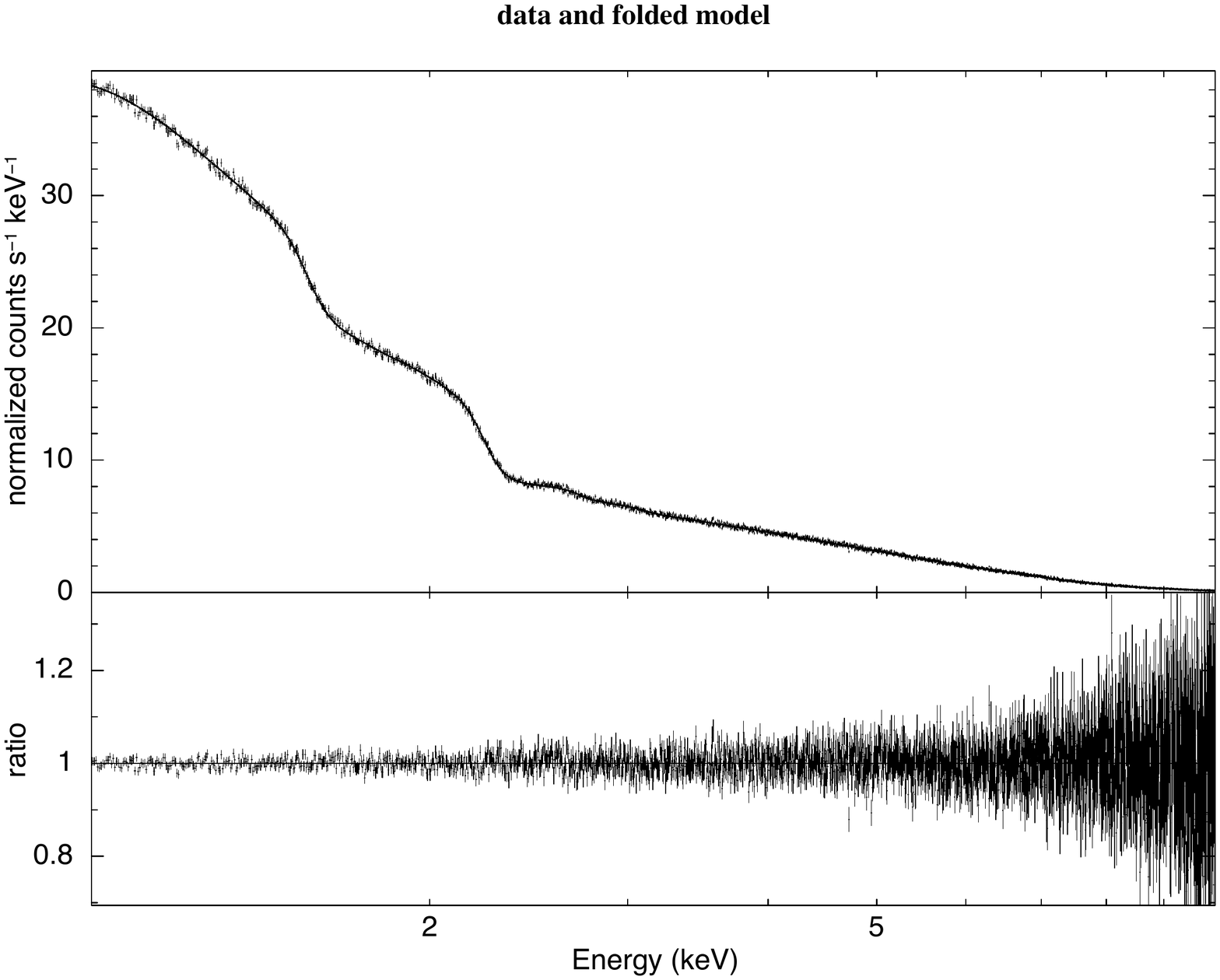}
\end{center}
\vspace{-0.5cm}
\caption{Best-fits for Configuration~III (top panels), Configuration~IV (central panels), and Configuration~V (bottom panels) from the simulations with XIS/Suzaku. The intensity profile of the reflection component in these simulations is described by the power-law $1/r^3$ (left panels) and the lamppost-inspired model $h/(h^2 + r^2)^{3/2}$ with $h = 2$ (right panels). In every panel, the top quadrant shows the simulated data and the Kerr-model best-fit (folded spectrum), while the bottom quadrant shows the ratio between the simulated data and the Kerr-model best-fit. See the text for more details. \label{f-suzaku} }
\end{figure*}

\begin{figure*}[h!]
\begin{center}
\includegraphics[type=pdf,ext=.pdf,read=.pdf,width=9cm]{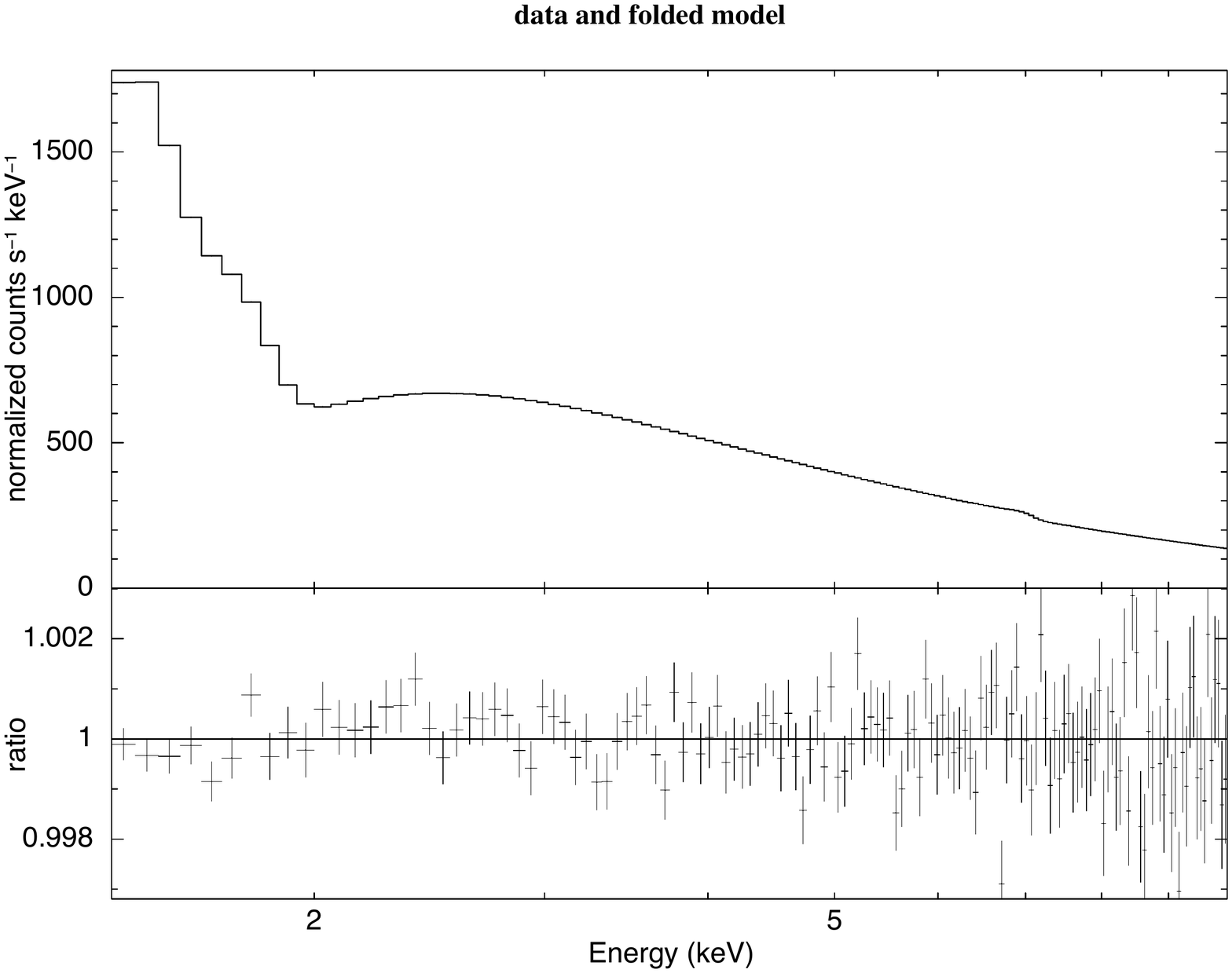}
\hspace{-0.5cm}
\includegraphics[type=pdf,ext=.pdf,read=.pdf,width=9cm]{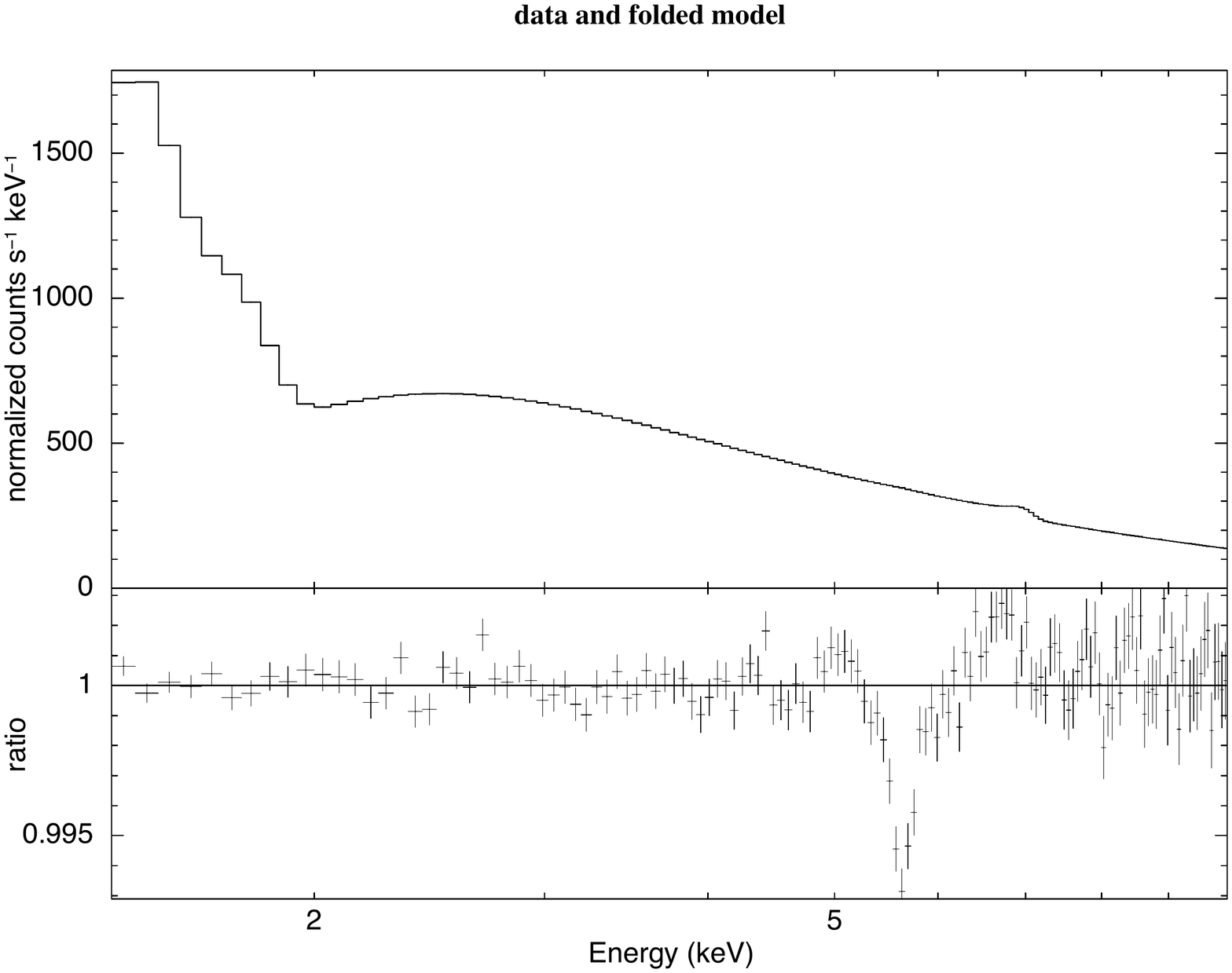} \\
\includegraphics[type=pdf,ext=.pdf,read=.pdf,width=9cm]{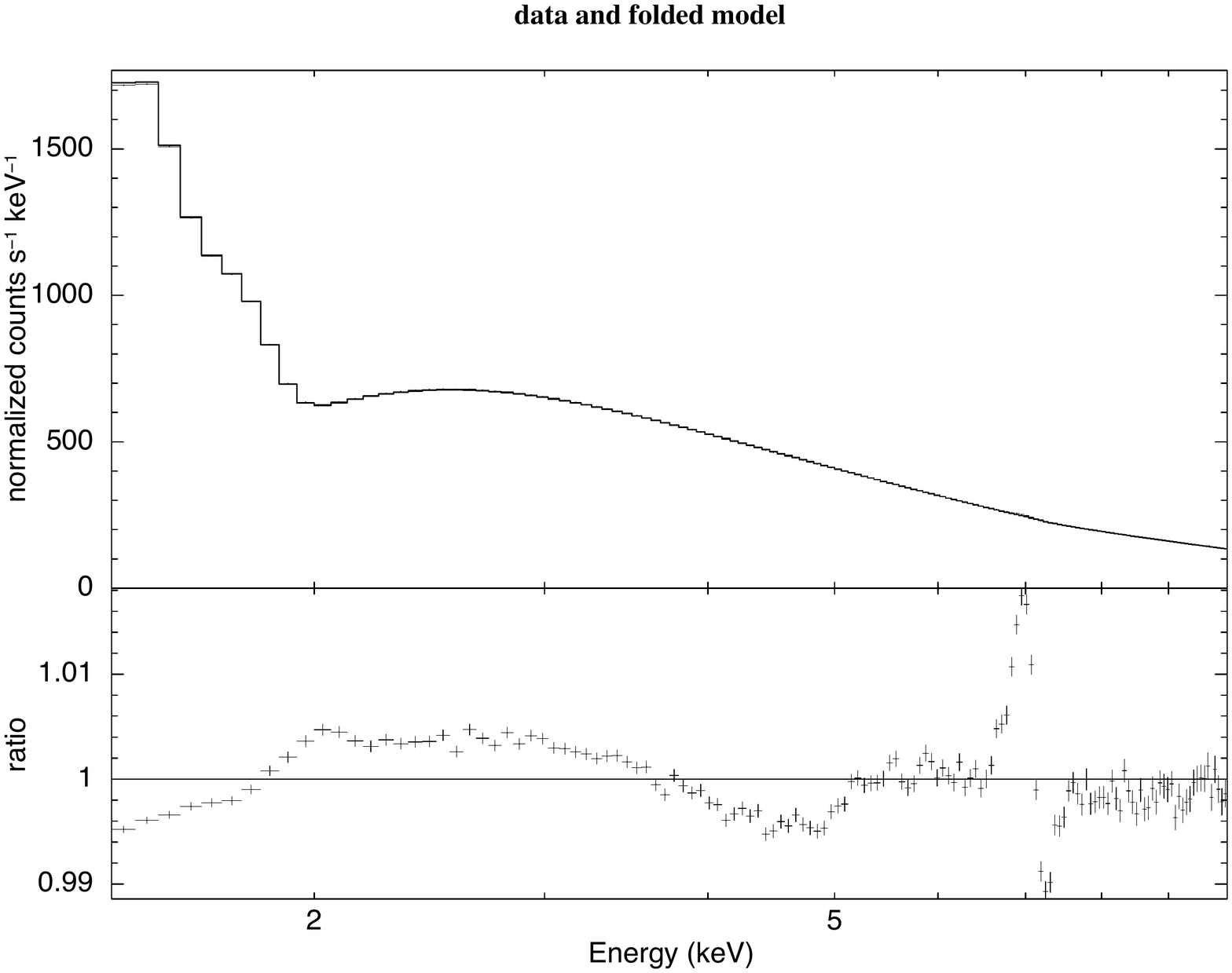}
\hspace{-0.5cm}
\includegraphics[type=pdf,ext=.pdf,read=.pdf,width=9cm]{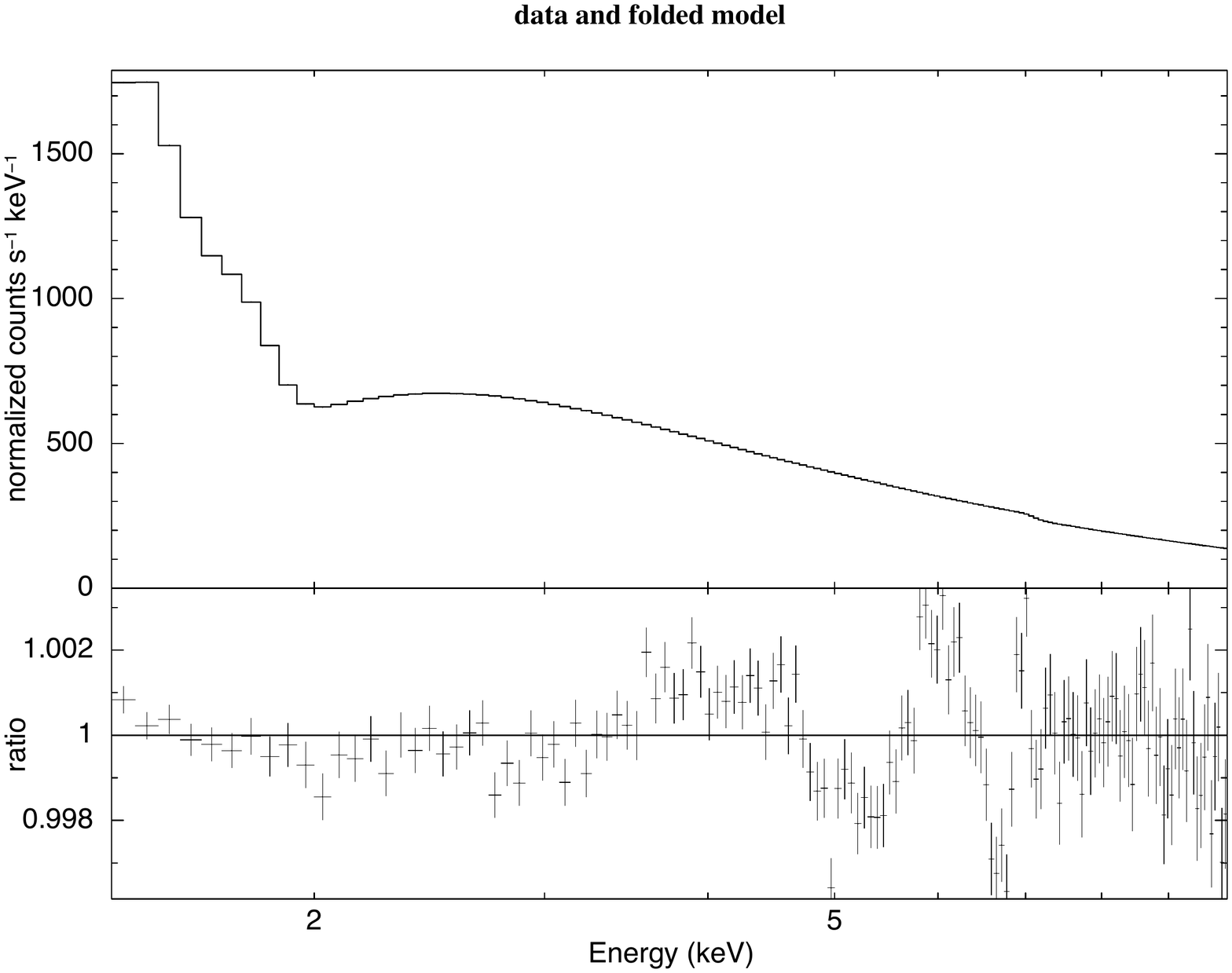} \\
\includegraphics[type=pdf,ext=.pdf,read=.pdf,width=9cm]{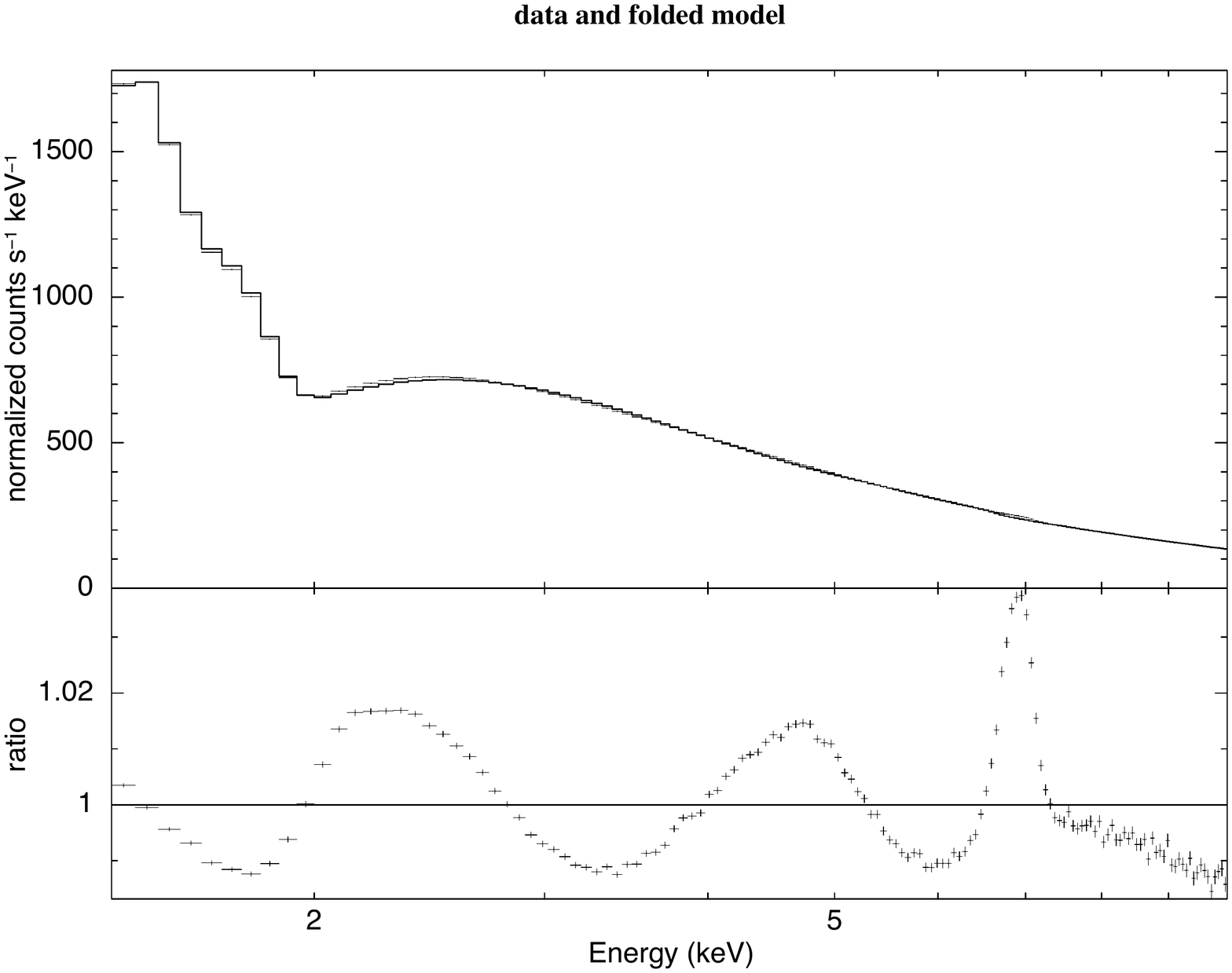}
\hspace{-0.5cm}
\includegraphics[type=pdf,ext=.pdf,read=.pdf,width=9cm]{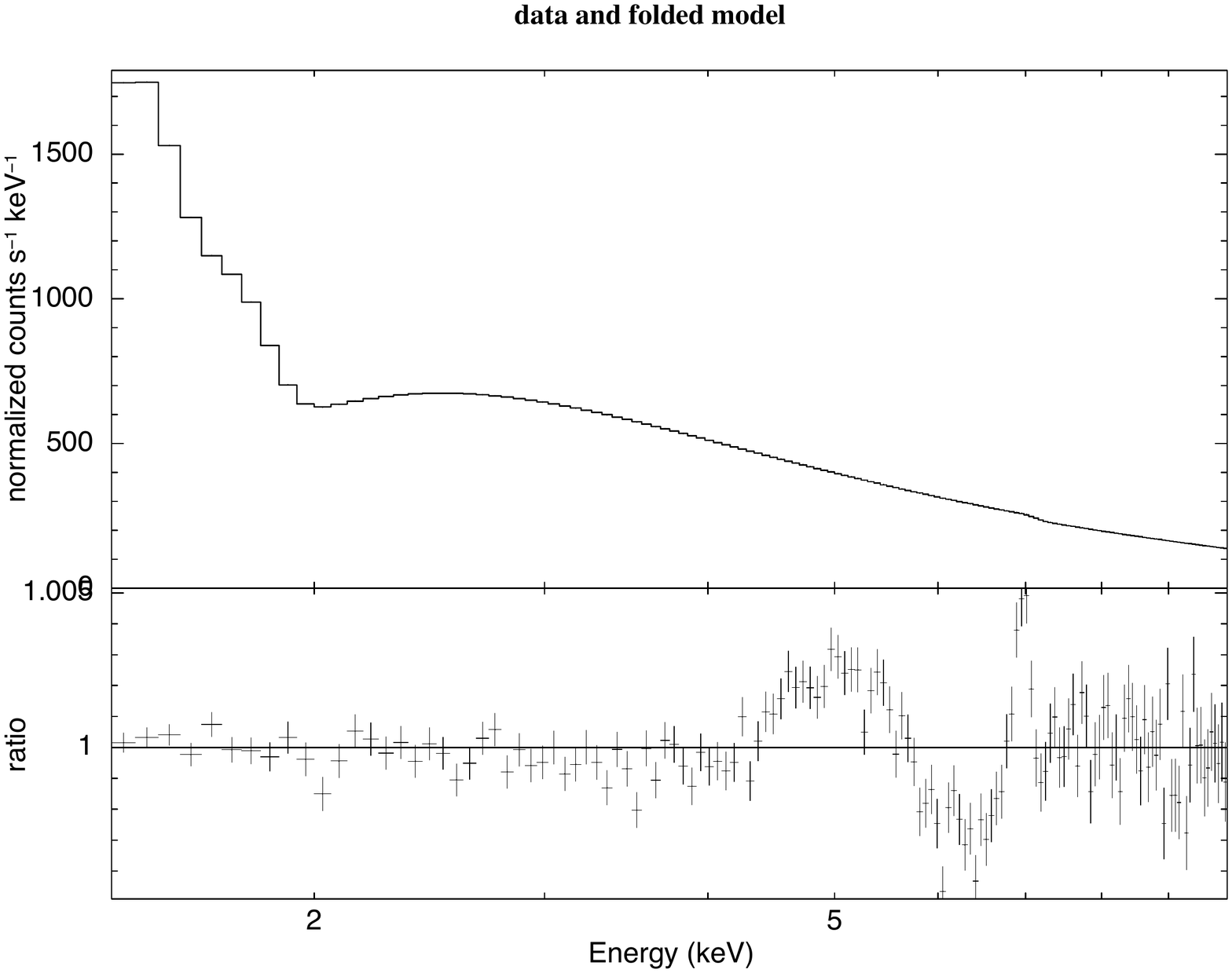}
\end{center}
\vspace{-0.5cm}
\caption{As in Fig.~\ref{f-suzaku} for the simulations with LAD/eXTP. \label{f-lad} }
\end{figure*}

Fig.~\ref{f-suzaku} shows our results for XIS/Suzaku, while Fig.~\ref{f-lad} is for LAD/eXTP. In each figure, the left panels are for the power-law profile $1/r^3$ and the right panels are for the lamppost-inspired profile. The top panels shows the results for Configuration~III, the central panels those for Configuration~IV, and the bottom panels are for Configuration~V. In every panel, the top quadrant shows the folded spectrum of the simulated data and the best-fit, while to bottom quadrant shows the ratio between the data and the best fit. The larger relative errors on the photon count at higher energies (particularly evident in Fig.~\ref{f-suzaku}) is due to the lower photon count (Poisson noise) at high energies. In the case of XIS/Suzaku, the effective area of the instrument drops above 7~keV and therefore the ratio between the model and the best fit becomes much noisier. Note that in Figs.~\ref{f-suzaku} and~\ref{f-lad} the $y$ axis of the top quadrant in each panel is in units of counts/s/keV and therefore it cannot be directly compared with the values of the $y$ axis in Fig.~\ref{f-iron}, where the flux is normalized. The word ``normalized'' in the $y$ axis in Figs.~\ref{f-suzaku} and~\ref{f-lad} refers to properties of the response file that can be ignored here (more details can be found in the Xspec manual).

The summary of the results with the best-fits is reported in Tab.~\ref{tab2} for XIS/Suzaku and in Tab.~\ref{tab3} for LAD/eXTP. The uncertainties on the parameters have to be taken with some caution, because they are not very relevant in this study for our conclusions and therefore they have not been estimated accurately. 


\begin{table*}[t]
\begin{tabular}{|c|cccccccc|}
\hline
\hspace{0.2cm} Configuration \hspace{0.2cm} & \hspace{0.2cm} Profile \hspace{0.2cm} &\hspace{0.2cm} $\chi^2_{\rm min,red}$ \hspace{0.2cm} & \hspace{0.1cm} $a_*$ \hspace{0.1cm} & \hspace{0.3cm} $i$ \hspace{0.3cm} & \hspace{0.1cm} $q_1$ \hspace{0.1cm} & \hspace{0.1cm} $q_2$ \hspace{0.1cm} & \hspace{0.3cm} $r_{\rm br}$ \hspace{0.3cm} & \hspace{0.3cm} $r_{\rm out}$ \hspace{0.3cm} \\
\hline 
III  & PL & 1.06 & 0.91(1) & 45(1) & 7(1) & 2.4(4) & 4.3(5) & 156(66) \\
IV & PL & 1.04 & $>0.99$ & 57(2) & 8.4(4) & -- & -- & -- \\
V & PL & 1.09 & 0.974(2) & 21(1) & 9.6(2) & -- & -- & -- \\
\hline
III & LP & 1.04 & 0.96(15) & 45.5(5) & 2.1 & -- & -- & -- \\
IV & LP & 0.98 & 0.96(1) & 46.7(8) & 3.7(1) & -- & -- & -- \\
V & LP & 1.04 & $>0.99$ & 46(1) & 3.7(3) & -- & -- & -- \\
\hline
\end{tabular}
\vspace{0.2cm}
\caption{Summary of the 100~ks simulations with XIS/Suzaku. In the second column ``profile'', PL stands for power-law (profile $\propto 1/r^3$) and LP stands for lamppost [profile $\propto h/(h^2 + r^2)^{3/2}$ with $h=2$]. The third column is the value of the reduced $\chi^2$ of the best-fit. From the fourth to the nineth columns, we show the best-fit values of the spin parameter ($a_*$), the viewing angle ($i$), the inner emissivity index ($q_1$), the outer emissivity index ($q_2$), the breaking radius in gravitational radii ($r_{\rm br}$), and the outer edge of the emission region in the accretion disk in gravitational radii ($r_{\rm out}$). When it is not possible to constrain the parameter, there is --. The uncertainties on the estimate of the parameters are the values in the round brackets and have to be taken only as a general guide. See the text for more details. \label{tab2} }
\vspace{0.8cm}
\begin{tabular}{|c|cccccccc|}
\hline
\hspace{0.2cm} Configuration \hspace{0.2cm} & \hspace{0.2cm} Profile \hspace{0.2cm} &\hspace{0.2cm} $\chi^2_{\rm min,red}$ \hspace{0.2cm} & \hspace{0.1cm} $a_*$ \hspace{0.1cm} & \hspace{0.3cm} $i$ \hspace{0.3cm} & \hspace{0.1cm} $q_1$ \hspace{0.1cm} & \hspace{0.1cm} $q_2$ \hspace{0.1cm} & \hspace{0.3cm} $r_{\rm br}$ \hspace{0.3cm} & \hspace{0.3cm} $r_{\rm out}$ \hspace{0.3cm} \\
\hline 
III  & PL & 1.15 & 0.931(2) & 44.83(6) & 3.98(9) & 3.28(5) & 4.2(3) & 104(23) \\
IV & PL & 31 & $>0.99$ & 59.1(3) & 7.82(6) & 4 & -- & 3.24(7) \\
V & PL & 257 & $>0.99$ & 31.5(2) & 10 & 3.98(3) & 3.02(2) & 20.7(6) \\
\hline
III & LP & 3.43 & 0.923(4) & 45.39(4) & 10 & 2.12(2) & 2.7(1) & 58.4(9) \\
IV & LP & 3.01 & 0.895(2) & 45.59(9) & 3.78(2) & 3.7(4) & -- & -- \\
V & LP & 4.02 & 0.989(4) & 45.67(8) & 8 & 3.67(3) & -- & 27(4) \\
\hline
\end{tabular}
\vspace{0.2cm}
\caption{As in Tab.~\ref{tab2} for the simulations with LAD/eXTP. See the text for more details. \label{tab3}}
\end{table*}

\section{Discussion \label{s-dis}}

The interpretation of the simulations with XIS/Suzaku are quite straightforward. The reduced $\chi^2$ of the best-fit is close to 1 and the Kerr model can well fit the data. The ratio between the simulated data and the best-fit in the bottom quadrant of each panel in Fig.~\ref{f-suzaku} is always close to 1 and we do not see unresolved features. The conclusion is that current X-ray missions cannot distinguish configurations~III, IV, and V from a Kerr BH. Note that we have considered a quite bright source (flux in the energy range 1-10~keV of $10^{-9}$~erg/s/cm$^2$) and a relatively long exposure time (100~ks). Moreover, the spectrum is a simple power-law with an iron line. These should be quite favorable ingredients for constraining the metric, and the fact that the Kerr model is capable of providing a good fit does not leave much space to other conclusions.

In the case of the $1/r^3$ intensity profile, the iron line shape are those shown in the left panel of Fig.~\ref{f-iron}. Those of Configuration~IV and, in particular, of Configuration~V are definitively different from those expected from the line computed in the Kerr metric with the $1/r^3$ intensity profile (see, for instance, Fig.~9 in Ref.~\cite{iron3}). Nevertheless, it is possible to find a good fit with a Kerr model employing a large value for the inner emissivity index. The latter indeed has the effect to increase the relative contribution from the radiation emitted from the inner part of the disk, which is strongly redshifted, and therefore it has the capability of reproducing the peak in the left panel of Fig.~\ref{f-iron} for the iron shape of Configuration~V (see again Fig.~9 in Ref.~\cite{iron3}). On the contrary, if we ``knew'' that the actual intensity profile goes like $1/r^3$, the fit would be horrible. In other words, there is a degeneracy between the intensity profile and the background metric, which could be fixed only in the case of a good understanding of the former (which, in turn, is determined by the geometry of the corona).

\begin{figure*}[h!]
\begin{center}
\includegraphics[type=pdf,ext=.pdf,read=.pdf,width=9cm]{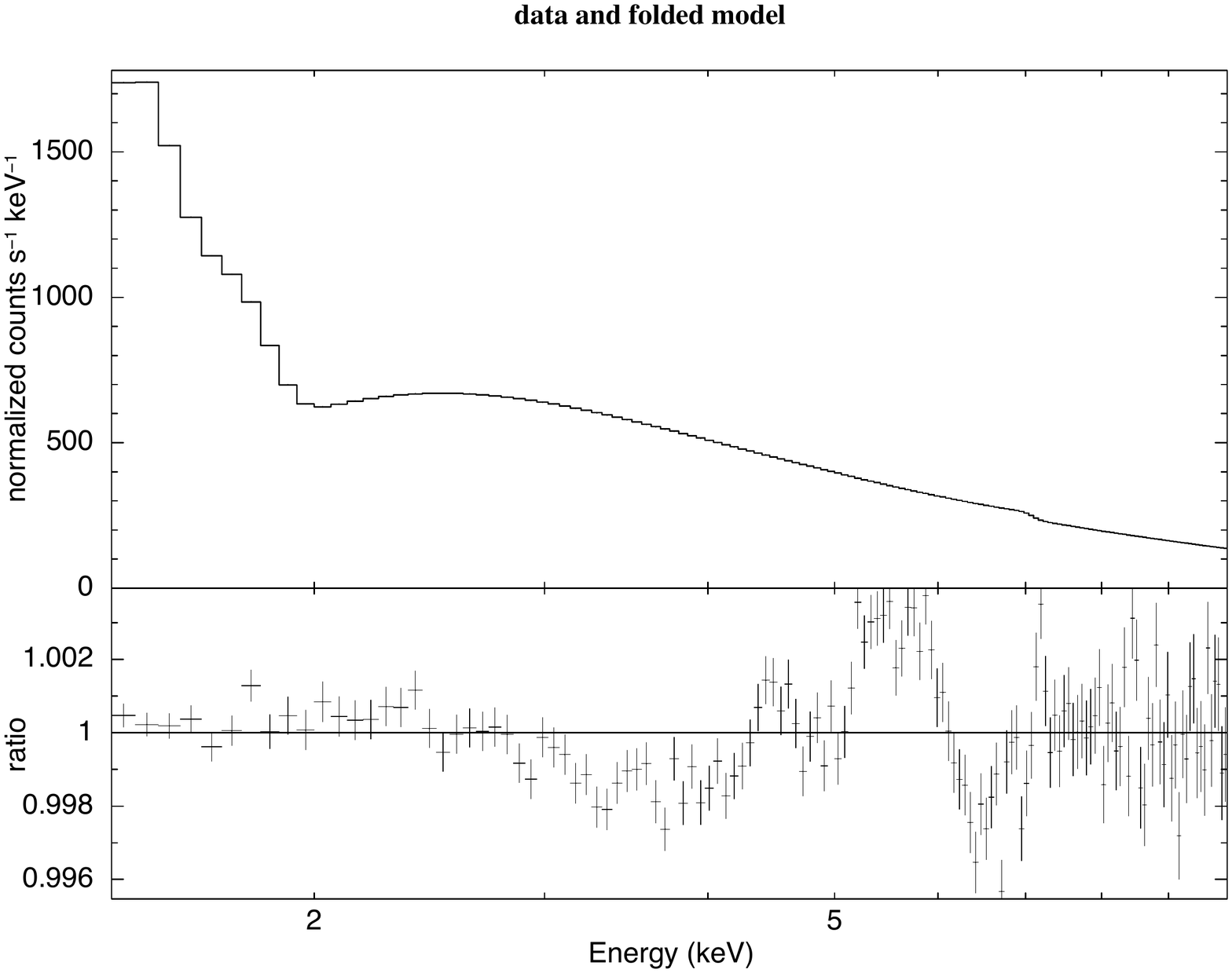}
\end{center}
\vspace{-0.5cm}
\caption{As in the top left panel of Fig.~\ref{f-lad}, but with the emissivity index $q$ of the intensity profile in the fit frozen to its correct value 3. \label{f-lad-bis} }
\end{figure*}

Let us now move to the simulations with LAD/eXTP. When we assume the $1/r^3$ intensity profile, the spectra of Configuration~IV and Configuration~V cannot be fitted with a Kerr model. The minimum of the reduced $\chi^2$ is a large number and there are no ambiguities. In the case of Configuration~III, the fit is good and the estimate of the parameters is not far from their actual value. In other words, Configuration~III is too similar to the spacetime of a Kerr BH and the impact of its Proca hair is too weak. If we freeze the emissivity profile to its exact form, we find that the Kerr model cannot provide a good fit, see Fig.~\ref{f-lad-bis}. The reduced $\chi^2$ of the best-fit is about 4.2 and we see unresolved features in the ratio between the simulated data and the best-fit. Once again, this shows that accurate tests of the Kerr metric using the iron line method will only be possible when the geometry of the corona is known and we have a theoretical prediction of the intensity profile of the reflection spectrum. Phenomenological models, like a power-law or a broken power-law often used with current data, are not adequate in the presence of high quality data possible with the next generation of X-ray facilities.

Last, we have the simulations with LAD/eXTP with the lamppost-inspired intensity profile. In this case, the iron lines can better mimic those expected in a Kerr metric assuming an intensity profile described by a broken power-law. Again, this can be easily interpreted with the fact that stringent tests of the metric around BHs will only be possible if we can predict theoretically the intensity profile from the geometry of the corona. In all the three spacetimes, the best-fit is bad ($\chi^2_{\rm min,red} > 3$) and we clearly see unresolved features in the quadrants showing the ratio between the simulated data and the best-fit models. However, we cannot exclude that a certain intensity profile can do it. In other words, it is probably possible to test the metric, but it is difficult to be able to get a conclusive result.

\section{Concluding remarks \label{s-con}}

In this paper, we have extended previous work to explore the capabilities of present and future X-ray missions to test astrophysical BH candidates using X-ray reflection spectroscopy, the so-called iron line method, and in particular the possibility of distinguishing Kerr BHs from Kerr BHs with Proca hair. The latter are non-vacuum solutions of 4-dimensional EinsteinÕs gravity, in which matter is described by a Proca (massive spin-1) field. The solutions can evade the no-hair theorem because of a time-dependence in the field, even if its energy-momentum tensor and the background metric are stationary.

We have simulated observations with XIS/Suzaku, to study the capabilities of present X-ray missions, and LAD/eXTP, to understand the opportunities with the next generations of X-ray facilities. We have simulated observations of a small set of Kerr BHs with Proca hair assuming the parameters for a bright BH binary, because the latter is likely the kind of source suitable for these tests. The technique can also be applied to supermassive BHs, but the constraints would be weaker because their luminosity is lower and the Poisson noise is thus higher.

The results of our simulations are shown in Fig.~\ref{f-suzaku} (XIS/Suzaku) and Fig.~\ref{f-lad} (LAD/eXTP) and summarized in Tab.~\ref{tab2} (XIS/Suzaku) and Tab.~\ref{tab3} (LAD/eXTP). In the case of XIS/Suzaku, it is not possible to distinguish these solutions from Kerr BHs: the fit is good, with $\chi^2_{\rm red,min}$ close to 1 and no unresolved features. This is true even for Configuration~V with the $1/r^3$ intensity profile, because it may be interpreted as a Kerr BH with a high value of the inner emissivity index.  We remark that our results do not exclude that very, or extremely, hairy BHs with Proca hair may exist whose reflection spectrum cannot be mimicked by any intensity profile on the Kerr background. To test this a more complete scanning of the parameter space must be performed. But it is quite instructive to realize how a BH which is considerably different from the standard Kerr solution can yield a similar phenomenology, as observed by current instruments. The situation is very different in the case of LAD/eXTP, where usually the simulations cannot provide good fits with a Kerr model. Configuration~III with the $1/r^3$ intensity profile can still provide an acceptable fit.

All these results have to be taken with caution, because of the numerous simplifications in our analysis and the limited sample studied. However, as a proof of concept, they illustrate important features, such as the degeneracy between the metric and  the intensity profile, even for extremely non-Kerr like metrics. As already stressed in Ref.~\cite{work3}, we would get a significant improvement in these tests if we could know the behavior of the intensity profile, because of this degeneracy. Moreover, the difference between current and future X-ray missions is clear.


\begin{acknowledgments}
The work of M.Z. and C.B. was supported by the NSFC (Grant No.~U1531117) and Fudan University (Grant No.~IDH1512060). C.B. also acknowledges the support from the Alexander von Humboldt Foundation. C.A.R.H. and E.R. acknowledge funding from the FCT-IF programme. This work was partially supported by the H2020-MSCA-RISE-2015 Grant No. StronGrHEP-690904, and by the CIDMA project UID/MAT/04106/2013.
\end{acknowledgments}


\end{document}